\documentclass[useAMS,usenatbib,usegraphicx]{mn2e}

\newcommand{\pin}{{\sc pinocchio}}
\newcommand{\gal}{{\sc morgana}}
\newcommand{\gs}{{\sc grasil}}
\newcommand{\be}{\begin{equation}}
\newcommand{\ee}{\end{equation}}
\newcommand{\bea}{\begin{eqnarray}}
\newcommand{\eea}{\end{eqnarray}}
\newcommand{\msunyr}{M$_\odot$ yr$^{-1}$}
\newcommand{\msun}{M$_\odot$}
\newcommand{\kms}{km s$^{-1}$}

\title[Assembly of massive galaxies] {Reproducing the assembly of
  massive galaxies within the hierarchical cosmogony}
\author[Fontanot et al.]{Fabio Fontanot$^{1,2}$, Pierluigi Monaco$^{2,3}$, Laura Silva$^3$, Andrea Grazian$^4$ \\
  $^1$Max-Planck-Institute for Astronomy, K\"onigstuhl 17, D-69117, Heidelberg, Germany \\
  $^2$Dipartimento di Astronomia, Universit\`a di Trieste, via Tiepolo 11, 34131 Trieste, Italy \\
  $^3$INAF-Osservatorio Astronomico di Trieste, via Tiepolo 11, 34131 Trieste, Italy \\
  $^4$INAF-Osservatorio Astronomico di Roma, via Frascati 33, I-00040Monteporzio, Italy \\
  email: fontanot@mpia-hd.mpg.de; monaco, silva@oats.inaf.it; grazian@mporzio.astro.it}

\begin{document}

\date{Accepted ... Received ...}

\pagerange{\pageref{firstpage}--\pageref{lastpage}} \pubyear{}

\maketitle
\label{firstpage}

\begin{abstract}
  In order to gain insight into the physical mechanisms leading to the
  formation of stars and their assembly in galaxies, we compare the
  predictions of the MOdel for the Rise of GAlaxies aNd Active nuclei
  ({\gal}) to the properties of $K$- and 850$\mu$-selected galaxies
  (such as number counts, redshift distributions and luminosity
  functions) by combining {\gal} with the spectrophotometric model
  {\gs}. We find that it is possible to reproduce the $K$- and $850
  \mu$-band datasets at the same time and with a standard Salpeter
  IMF, and ascribe this success to our improved modeling of cooling in
  DM halos.  We then predict that massively star-forming discs are
  common at $z\sim2$ and dominate the star-formation rate, but most of
  them merge with other galaxies within $\sim$100 Myr.  Our preferred
  model produces an overabundance of bright galaxies at $z<1$; this
  overabundance might be connected to the build-up of the diffuse
  stellar component in galaxy clusters, as suggested by Monaco et al.
  (2006), but a naive implementation of the mechanism suggested in
  that paper does not produce a sufficient slow-down of the evolution
  of these objects.  Moreover, our model over-predicts the number of
  $10^{10}-10^{11}$ {\msun} galaxies at $z\sim1$; this is a common
  behavior of theoretical models as shown by Fontana et al.  (2006).
  These findings show that, while the overall build-up of the stellar
  mass is correctly reproduced by galaxy formation models, the
  ``downsizing'' trend of galaxies is not fully reproduced yet. This
  hints to some missing feedback mechanism in order to reproduce at
  the same time the formation of both the massive and the small
  galaxies.
\end{abstract}

\begin{keywords}
galaxies: formation -- galaxies: evolution
\end{keywords}

\section{Introduction}
\label{section:introduction}
The Lambda Cold Dark Matter ($\Lambda$CDM) cosmology is consistent
with a large body of observations of the large-scale Universe (see,
e.g., Spergel et al. 2006).  Then, the predicted hierarchical
evolution of Dark Matter (DM) perturbations subject to gravitational
instability provides a standard framework to study the formation and
evolution of luminous structures in the Universe.  However, while the
cosmological framework is fixed with a small uncertainty, several open
questions, regarding to the formation and evolution of galaxies, arise
from the complex evolution of baryons within the potential wells of
the DM halos.

Galaxy formation is observationally constrained by many multi
wavelength surveys of deep fields (e.g. COMBO17, Wolf et al., 2001;
DEEP2, Davis et al., 2003; GOODS, Giavalisco et al.  2004; GEMS, Rix et
al. 2004; UKIDSS, Lawrence et al., 2006; COSMOS, Scoville et al.,
2007; etc.).  To compare with these datasets, many thousands of
galaxies must be generated by models.  This makes a straightforward
numerical approach problematic (it has been attempted, e.g, by
Nagamine et al.  2005, Saro et al., 2006, Robertson et al.  2007), so
that simpler and quicker models have been developed, based on sets of
recipes that address the various processes involved.  These
``semi-analytical'' models (see e.g. Somerville et al. 2001, 2004;
Granato et al.  2004; Menci et al., 2004; Baugh et al. 2005; Kang et
al., 2005; Bower et al., 2006; Croton et al.  2006; Cattaneo et al.,
2006; De Lucia et al., 2006; Monaco, Fontanot \& Taffoni, 2007) have
been tested against an impressive number of observational constraints.
In this decade-long testing process, many specific models have been
unsuccessful in reproducing constraints like the high-mass cutoff of
the luminosity function (Benson et al. 2003), the level of
$\alpha$-enhancement in elliptical galaxies (Thomas 2005; Nagashima et
al., 2005), the redshift distribution of K-band sources (Cimatti et
al. 2002a), the surface density of EROs (Cimatti et al.  2002b, Daddi
et al.  2002; Smith et al. 2001).

Many of these difficulties have been overcome by later versions of the
models, but some of them have required strong assumptions: for
instance, a top-heavy IMF in starbursts is required by Baugh et al.
(2005) to reproduce the sub-mm counts.  In the most recent models (see
i.e.  Bower et al.  2006; Croton et al., 2006) the cutoff of the
galaxy luminosity function and the bimodality of galactic colors is
obtained only if cooling flows in large halos are quenched by AGN
feedback, i.e.  by the energy emerging from massive black holes
accreting at a relatively low rate (in the so-called {\it radio
  mode}).

A recently highlighted observational trend of galaxy formation is the
so-called {\it downsizing} of galaxies: at variance with the
hierarchical trend of DM halos, more massive galaxies tend to form
their stars earlier and in a shorter period than smaller galaxies,
which experience more prolonged star-formation histories.  While this
general trend is known to be not incompatible with cosmology once
stellar and AGN feedback are properly taken into account (see, e.g.,
Granato et al., 2004; Neisteinn, van den Bosch \& Dekel, 2006), recent
observations show that the details, like for instance the (almost)
parallel evolution of the star formation density for galaxies of
different mass (Zheng et al. 2007), are still not reproduced.

Massive elliptical galaxies have always been remarkably elusive
objects in this regard.  The first versions of hierarchical galaxy
formation models (see, e.g.  White 1996) predicted that these galaxies
form late ($z\sim1$) by the merging of already assembled discs, while
evidence from stellar populations (see Matteucci 1996 for a review),
the tightness of the fundamental plane (Renzini \& Ciotti, 1993), the
evolution of the color-magnitude relation (Kodama et al., 1998;
Blakeslee et al., 2003; Ellis et al. 2006) and the local $Mg_2-\sigma$
relation (Bernardi et al.  2003) suggested that they formed early
($z>2$) in a short burst of star formation.  Clearly, in a
hierarchical Universe the age of stars does not need to coincide with
the assembly age of the galaxy, defined as the time at which the most
massive progenitor has at least half of the final stellar mass.
Massive ellipticals could then be assembled late by {\it dry mergers}
of other ellipticals, so as to preserve the oldness of their stellar
populations while producing a low assembly redshift (De Lucia et al.
2006).  This possibility is severely constrained by the modest (if
any) evolution of the high end of the stellar mass function since
$z=1$ (Cimatti, Daddi \& Renzini, 2006): while models predict a
doubling of stellar masses (De Lucia \& Blaizot, 2007), evidence
excludes an evolution larger than $\sim$0.2 dex (as roughly estimated
by Monaco et al. 2006; see the references therein).  Recently, Monaco
et al.  (2006; see also Conroy, Weschler, \& Kravtsov, 2007) have
proposed that the formation of a diffuse stellar component in galaxy
clusters by scattering of stars during dry mergers (see, e.g., Murante
et al. 2007) may conspire to decrease the expected evolution of the
high end of the stellar mass function.

The early formation of massive galaxies and their following (almost
passive) evolution are best constrained by deep observations in the
sub-mm band, suited to reveal obscured star formation events at high
redshift, and in the $K$-band, suited to probe stellar masses at lower
redshift.

The warm
dust present in star forming clouds absorbs most of the UV/blue
photons emitted by young stars and reprocesses them to the FIR,
becoming the dominant contributor in that band.  Moreover, the steeply
decreasing shape of the galactic SEDs from $\sim$100$\mu$m to $\sim$1
mm gives a negative K-correction that promotes the observation of
starbursts in the sub-mm bands up to $z\sim5$. Therefore strong
starbursts at high redshift are more easily observed in the sub-mm
than in the optical.  The sub-mm emission is measurable in a few
windows, most notably that at 850 $\mu$m.  Observations with the
Submillimiter Common-User Bolometer Array (SCUBA) on the James Clerk
Maxwell Telescope have highlighted the presence of a population of
high-redshift massive starbursts (see, e.g., Smail et al. 1997; Hughes
at al.  1998; Chapman et al. 2002; Scott et al.  2002)
%
commonly interpreted as galaxies forming stars at rates of hundreds if
not thousands of \msunyr.  Given the poor angular resolution of SCUBA
images, the identification of optical counterparts is difficult, and
can be achieved using interferometric images at longer wavelengths.
The resulting redshift distribution is thought to peak at $z\sim2.4$
(Chapman et al. 2003, 2005).

The $K$-band is a very good tracer of the stellar mass at $z\la1.5$
(Gavazzi et al. 1996), it is almost unaffected by dust extinction, and
requires small $K$-corrections that weakly depend on the morphological
type, so it is ideal to follow the assembly of the bulk of stellar
mass.
%
The local $K$-band LF has been measured with great accuracy by the
2MASS collaboration (Cole et al.  2001; Kochanek et al. 2001).  In
this paper we will focus on four $K$-band surveys with (photometric or
spectroscopic) redshift coverage. (i) The K20 survey (Cimatti et al.
2002a, Pozzetti et al. 2003) is a $K<20$ (corresponding to
$K_{AB}<21.84$) limited sample, covering $52 arcmin^2$, with a very
high redshift completeness ($> 90\%$).  (ii) The GOODS-MUSIC (Grazian
et al. 2006) catalogue is a multicolor sample extracted from the deep
and wide survey conducted over the Chandra Deep Field South in the
framework of the GOODS project. Here we use the $K_{AB}<23.5$ limited
galaxy sample defined in Fontana et al.  (2006).  This sample covers
$143.2 arcmin^2$; $28 \%$ of the galaxies have a spectroscopic
redshifts, used to train photometric redshift estimates for all other
galaxies.  (iii) The UKIDSS Ultra Deep Survey (UDS, Dye et al., 2006)
galaxy sample is a complete catalog of $K_{AB}<22.5$ selected galaxies
over $0.6 deg^2$; for each object in the sample a photometric estimate
of the redshift is provided. (iv) The VIMOS-VLT Deep Survey (VVDS) (Le
Fevre et al., 2005) is a spectroscopic survey designed to measure
redshift for $\sim 10^5$ sources selected, nearly randomly from a
photometric catalogue. Here we consider a $K$-selected sample, with
either photometric and spectroscopic redshifts obtained by Pozzetti et
al. (2007) combining a $K_{AB}<22.34$ limited sample ($20\%$ redshift
completeness) defined over a $442 arcmin^2$ area, with a
$K_{AB}<22.84$ limited sample ($29\%$ redshift completeness) defined
over a $172 arcmin^2$ area.

This paper is the third of a series devoted to describe the MOdel for
the Rise of GAlaxies aNd Active nuclei ({\gal}).  With respect to
similar models of galaxy formation, {\gal} presents a different, more
sophisticated treatment of the mass and energy flows between galactic
phases (cold and hot gas, stars) and components (bulge, disc, halo).
In particular, the process of radiative cooling of the shocked gas is
treated with a new model (tested against simulations in Viola et al.,
in preparation), while feedback is inserted following the model by
Monaco (2004) and galaxy winds and super-winds are allowed.  The model
is described in detail in Monaco, Fontanot \& Taffoni (2007; hereafter
paper I).  The prediction of the properties of the AGN population is
presented in Fontanot et al. (2006; hereafter paper II), and the
prediction of the evolution of the stellar mass function, together
with those of other similar models, has been compared to the results
inferred from the GOODS-MUSIC data in Fontana et al. (2006).  As
mentioned above, {\gal} has been used by Monaco et al. (2006) to
address the lack of evolution of the high end of the stellar mass
function and its connection with the building of the diffuse stellar
component in galaxy clusters.  In this paper we compare the
predictions of {\gal} to the data mentioned above of deep fields in
sub-mm (850 $\mu$m) and $K$ bands to test to what extent the model is
able to reproduce the formation and assembly of massive galaxies.  To
this aim, we have combined {\gal} with the spectrophotometric code
{\gs} (Silva et al.  1998) that computes the UV to radio SEDs of model
galaxies, including a three-dimensional bulge+disk geometry with a
two-phase interstellar medium, the radiative transfer through the
dusty ISM, a realistic dust grain model, and a direct computation of
the dust temperature distribution.

The paper is organized as follows. In section \ref{section:morgana} we
describe the main properties of the {\gal} model. In section
\ref{section:fontanot} we describe how we compute luminosity
functions, number counts and redshift distributions interfacing the
output of the model with {\gs}.  In section \ref{section:results} we
present our results, a discussion is given in
Section~\ref{section:discussion}, and in
Section~\ref{section:conclusions} we give our conclusions.  Throughout
this work we assume, whenever necessary, the concordance cosmological
model $\Omega_\Lambda=0.7$, $\Omega_m=0.3$, $H_0=70 \, Km \, s^{-1} \,
Mpc^{-1}$, $\sigma_8=0.9$.

\section{MODEL}
\label{section:model}

\subsection{Galaxy formation model: {\gal}} 
\label{section:morgana}

{\gal} is
described in full detail in Monaco, Fontanot \& Taffoni (2007), while
AGN accretion is described in Fontanot et al. (2006). We give here
only a brief description, aimed at highlighting the main processes
included in the model.
\subsubsection{Algorithm} \label{section:algorithm}

{\gal} follows the typical scheme of semi-analytic models, with some
important differences.  Each DM halo contains one galaxy for each
progenitor\footnote{
Each DM halo forms through the merging of many halos of smaller mass,
called progenitors. At each merging the largest halo survives (it
retains its identity), the others become substructure of the largest
one.  The main progenitor is the one that survives all the mergings.
The mass resolution of the box used for computing the merger trees
sets the smallest progenitor mass, as explained in
section~\ref{section:runs}.
}; the galaxy associated with the main progenitor is the central galaxy.
Baryons in a DM halo are divided into three components, namely a halo, a bulge
and a disc. Each component contains three phases, namely cold gas, hot gas and
stars. For each component the code follows the evolution of its mass, metal
content, thermal energy of the hot phase and kinetic energy of the cold phase.

The main processes included in the model are the following.

(i) The merger trees of DM halos are obtained using the {\pin} tool
(Monaco et al. 2002; Monaco, Theuns \& Taffoni 2002; Taffoni, Monaco \& Theuns
2002).

(ii) After a merging of DM halos, dynamical friction, tidal stripping
and tidal shocks on the satellite (the smaller DM halo, with its
galaxy at the core) lead to a merger with the central galaxy or to
tidal destruction as described by Taffoni et al. (2003). At each
merger a fraction of the satellite stars is scattered to the stellar
halo component (Murante et al. 2007; Monaco et al. 2006).  These stars
are not associated to galaxies but to the intra-cluster light.

(iii) The evolution of the baryonic components is performed by numerically
integrating a system of equations for all the mass, energy and metal flows;
this allows not to be restricted to linear dynamics.

(iv) The intergalactic medium infalling on a DM halo is shock-heated, as well
as the hot halo component of merging satellites (which is given to the main
halo) and that of the main halo in case of major merger ($M_{\rm sat} > 0.2
\times M_{\rm tot}$). Following Wu, Fabian \& Nulsen (2001), shock-heating is
implemented by assigning to the infalling gas a specific thermal energy equal
to 1.2 times the specific virial energy, $-0.5\, U_{\rm H}/M_{\rm H}$ (where
$M_{\rm H}$ and $U_{\rm H}$ are the mass and binding energy of the DM halo).

(v) The profile of the hot halo gas is computed at each time-step by
solving the equation for hydrostatic equilibrium with a polytrophic
equation of state and an assumed polytrophic index $\gamma_p=1.2$.  No
hot gas is present within a cooling radius $r_{\rm cool}$, which is
set to a vanishingly small value at major mergers.

(vi) The cooling flow is computed by integrating the contribution to
radiative cooling of each spherical shell, taking into account the
heating from (stellar and AGN) feedback from galaxies.  Given the
importance of cooling for the results presented in this paper, we give
here some detail of the cooling model. If $T_{g0}$ and $\rho_{g0}$ are
the temperature and density of the hot halo gas extrapolated to $r=0$,
$\mu_{\rm hot}m_p$ its mean molecular weight and $r_s=r_{\rm H}/c_{\rm
  nfw}$ the scale radius of the halo (of radius $r_{\rm H}$ and
concentration $c_{\rm nfw}$), then the mass cooling flow $\dot{M}_{\rm
  co,H}$ results:

\be \dot{M}_{\rm co,H} = \frac{4 \pi r_s^3\rho_{g0}}{t_{\rm cool,0}}
\times {\cal I}(2/(\gamma_p-1))
\label{eq:coolingflow}
\ee

\noindent
where the integral ${\cal I}(\alpha)$ is defined as $\int_{r_{\rm
cool}/r_s}^{c_{\rm nfw}} \{ 1- a [1-\ln(1+t)/t ]\}^{\alpha}t^2 dt$,
with $a=[3T_{\rm vir}(\gamma_p-1)c_{\rm nfw}(1+c_{\rm nfw})]/
\{\gamma_p T_{g0}[(1+c_{\rm nfw})\ln(1+c_{\rm nfw})-c_{\rm nfw}]\}$
($T_{\rm vir}$ being the virial temperature of the halo).  The cooling
time $t_{\rm cool,0}$ is computed using the central density (the
density gradient is taken into account by the integral ${\cal I}$) and
the temperature at $r_{\rm cool}$ (the temperature gradient is
neglected):

\be t_{\rm cool,0} = \frac{3kT_g(r_{\rm cool}) \mu_{\rm hot}
  m_p}{2\rho_{g0} (\Lambda_{\rm cool}-\Gamma_{\rm heat})}
\label{eq:cool_heat}
\ee

\noindent
Here $\Lambda_{\rm cool}$ is the metal-dependent Sutherland \& Dopita (1993)
cooling function and the heating term $\Gamma_{\rm heat}$ is computed assuming
that the energy flow $\dot{E}_{\rm hw,H}$ fed back by the galaxy is given to
the cooling shell:

\be \Gamma_{\rm heat} = \frac{\dot{E}_{\rm hw,H}}{4\pi r_s^3 {\cal
    I}(2/(\gamma_p-1))} \left(\frac{\mu_{\rm
      hot}m_p}{\rho_{g0}}\right)^2
\label{eq:heating}
\ee

\noindent
Whenever $\Gamma_{\rm heat}>\Lambda_{\rm cool}$ the cooling flow is
quenched.  The cooling radius $r_{\rm cool}$ is treated as a dynamical
variable whose evolution takes into account the hot gas injected by
the central galaxy ($\dot{M}_{\rm hw,H}$):

\be \dot{r}_{\rm cool} = \frac{\dot{M}_{\rm co,H}-\dot{M}_{\rm hw,H}} {4\pi
  \rho_g(r_{\rm cool})r_{\rm cool}^2}\, .
\label{eq:drcool} \ee

\noindent
This equation is valid if pressure is balanced at $r_{\rm cool}$,
an assumption which clearly does not hold in general.  We mimic the
pressure force acting on mass shell at $r_{\rm cool}$ as follows:

\be
\dot{r}'_{\rm cool} = \dot{r}_{\rm cool} - c_s
\label{eq:drcool2}
\ee

\noindent
where $c_s$ is the sound speed computed at $r_{\rm cool}$. This recipe
is different from that used in most other semi-analytical models,
where a cooling radius is computed by inverting the cooling time as a
function of radius. Viola et al. (2007) have compared analytic cooling
models to N-body + hydro simulations showing that, while the
"classical" cooling model significantly underestimates the amount of
cooled mass, the present cooling model gives a very good fit.


(vii) When the hot halo phase is heated by feedback beyond the virial
temperature, it can leave the DM halo in a galactic super-wind.
%
%
A similar thing happens to the cold halo gas when it is accelerated by
stellar feedback.  To compute the time at which the ejected gas falls
back into a DM halo, its merger history is scrolled forward in time
until the circular velocity is larger than the (sound or kinetic)
velocity of the gas at the ejection time.

(viii) The cooling gas is let infall on the central galaxy on a
dynamical time-scale (computed at $r_{\rm cool}$).  It is divided
between disc and bulge according to the fraction of the disc that lies
within the half-mass radius of the bulge.


(ix) The gas infalling on the disc keeps its angular momentum; disc sizes
are computed with an extension of the Mo, Mao \& White (1998) model that
includes the contribution of the bulge to the disc rotation curve.

(x) Disc instabilities and major mergers of galaxies lead to the formation of
bulges.  We also take into account a possible disc instability driven by
feedback. In minor mergers the satellite mass is given to the bulge component
of the larger galaxy.

(xi) Star formation and feedback in bulges and discs are inserted following the
model of Monaco (2004). According to that model, the regime of stellar
feedback in a galaxy depends mainly on the density and vertical scale-length
of the galactic system. In thin systems, like spiral discs
with gas surface density $\Sigma_{\rm cold,D}$
and fraction of cold gas $f_{\rm cold,D}$, the timescale for star formation
$t_{\rm\star,D}$ is predicted to be:

\be t_{\rm\star,D} = 9.1\; \left(\frac{\Sigma_{\rm cold,D}}{1\ {\rm
      M}_\odot\ {\rm pc}^{-2}}\right)^{-0.73} \left(\frac{f_{\rm
      cold,D}}{0.1}\right)^{0.45}\ {\rm Gyr}
\label{eq:fbthin2}\ee

\noindent
Due to the correlation of $f_{\rm cold,D}$ and $\Sigma_{\rm cold,D}$ (galaxies
with higher gas surface density consume more gas), this relation is compatible
with the Schmidt law.
Thick systems like star-forming bulges (or mergers) are dominated by
transients which are very difficult to model, so
the straightforward Schmidt law is used:

\be t_{\rm \star,B} = 4\; \left(\frac{\Sigma_{\rm cold,B}}{1\ {\rm
      M}_\odot\ {\rm pc}^{-2}}\right)^{-0.4} \ {\rm Gyr}
\label{eq:schmidt} \ee

\noindent
In both cases, hot gas is ejected to the halo (in a hot galactic wind) at a
rate equal to the star-formation rate (as predicted by Monaco 2004), though
massive bulges with circular velocity $V_{\rm B}\ga300$ {\kms} are able to
bind their
hot phase component.  The thermal energy of this
re-heated gas is not scaled to the DM halo circular velocity but to the
energy of exploding SNe.

(xii) In star-forming bulges cold gas is ejected in a cold galactic wind by
kinetic feedback due to the predicted high level of turbulence driven by SNe
(this process is described in full detail in paper II,  Section~2.2).

(xiii) Accretion of gas onto massive black holes (starting from small
seeds present in all galaxies) is connected to the ability of cold gas
to loose angular momentum by some (unspecified) mechanism driven by
star formation.  This is explained in full detail in paper II.

(xiv) Because accretion onto black holes is triggered by star
formation, AGN feedback can quench cooling flows only after their
start.  Alternatively, the cooling flow can be quenched when a
fiducial energy criterion is met, without letting any gas fall to the
galaxy and form stars.  This ``forced quenching'' procedure has been
used in paper II and will be discussed in a forthcoming paper.

(xv) Metal enrichment is self-consistently modeled in the instantaneous
recycling approximation.

\subsubsection{Runs} \label{section:runs}

All models are based on the same {\pin} run we introduced in paper I, a
$512^3$ realization of a $150 Mpc$ comoving box ($h = 0.7$).  The mass
particles is $1.0\times 10^9$ \msun, and the smallest halo we consider
contains 50 particles, for a mass of $5.1\times 10^{10}$ \msun.  The branches
of the DM halo merger trees are tracked starting from a mass of 10 particles,
corresponding to $1.0\times10^{10}$ \msun; this is the mass of the smallest
progenitors. For sake of comparison, the corresponding values for the
Millennium Simulation (Springel et al.  2005) are $1.2 \times 10^{9}$ for the
particle mass and $2.5 \times 10^{10}$ for the smallest progenitor.  We have
tested the overall stability of our results by running the model over two
other $512^3$ boxes of size $200$ and $100$ Mpc (see paper I for more
details).

The stellar mass of the typical galaxy contained in the smallest DM
halo at $z=0$ is $3 \times 10^{8}$ {\msun}. In order to estimate the
completeness limit for the stellar mass function we consider an higher
resolution {\pin} run ($512^3$ realization of a $100 Mpc$ comoving
box, see also paper I, appendix B, for a discussion about the
numerical stability of the model). We study the typical stellar mass
of the galaxies belonging to $2.5-5.1\times 10^{10}$ {\msun} DM halos
and we estimate a completeness limit of $4.5\times 10^{8}$ {\msun}.
For each run we compute the evolution of (up to) 300 trees (i.e. DM
halos at $z=0$) per logarithmic bin of halo mass of width 0.5 dex.
This implies that while all the most massive halos are considered,
smaller halos are randomly sparse-sampled.  To properly reconstruct
the statistical properties of galaxies we assign to each tree a weight
$w_{\rm tree}$ equal to the inverse of the fraction of selected DM
halos in the mass bin.

A standard parameter choice was presented in paper I.  However, it was
noticed there that spiral discs tend to be more compact than in the
real Universe.  A Schmidt-Kennicutt-like law
(equation~\ref{eq:fbthin2}) is used to compute star formation rates,
so this results in higher surface densities, shorter star formation
timescales and lower gas fractions at $z=0$.  Forcing DM halo
concentrations to lower values allows to alleviate this problem, while
reproducing the zero-point of the Tully-Fisher relation and preserving
a good fit of the Schmidt-Kennicutt law.  We then scale concentrations
so as to take a given value for a $10^{12}$ {\msun} DM halo at $z=0$.
The value was proposed to be 4 in paper I; however, we noticed that
this choice tends to lower the stellar mass function at the knee, so
we use the slightly higher value of 6 as a good compromise.  Moreover,
following Monaco et al. (2006) we scatter to the halo stellar
component 40\% of the stellar mass of satellites at each galaxy
merger.  Finally, we implement AGN feedback by adopting the ``forced
quenching'' procedure of paper II; this choice has little influence on
the results presented here, and will be discussed in a forthcoming
paper.  Also, we have checked that the inclusion of quasar-triggered
galaxy winds, necessary to reproduce the accretion history of massive
black holes (paper II), does not change the qualitative behavior of
the models.

These changes of parameters do not influence in any way the
conclusions drawn in this paper.  In fact, we are not proposing a
``best fit'' model of the galaxy population; on the contrary, we will
show that {\gal} is able to reproduce many observables, but its
agreement with data breaks as soon as one tries to reproduce some
relevant aspects of the ``downsizing'' trend, so that a global
fine-tuning of the model is not possible.  The most important point of
such an investigation is to understand the limits of these models and
the origin of the disagreement with data.  Fine-tuning to fit specific
datasets, like the local luminosity functions, will be required in
other contexts, for instance to create galaxy mock catalogues.

\subsection{SED model: {\gs}} \label{section:fontanot}

%
%


For each galaxy modeled by {\gal} we compute the corresponding UV to
radio SED with {\gs}.  The details of the code are given in Silva et
al. (1998) (and the subsequent updates and improvements in Silva 1999;
Granato et al. 2000; Bressan, Silva, Granato 2002; Panuzzo et al.
2003; Vega et al. 2005), while we summarize here the main features.
(i) Stars and dust are distributed in a bulge (King profile) + disk
(radial and vertical exponential profiles) axisymmetric geometry. (ii)
We consider the clumping of both (young) stars and dust, through a
two-phase interstellar medium with dense giant molecular clouds (MCs)
embedded in a diffuse (``cirrus'') phase. (iii) The stars are assumed
to be born within the optically thick MCs and gradually to escape from
them as they get older on a time-scale $t_{\rm esc}$. This gives rise
to an age-selective extinction, with the youngest and most luminous
stars suffering larger dust extinction than older ones. (iv) The dust
consists of graphite and silicate grains with a distribution of grain
sizes, and Polycyclic Aromatic Hydrocarbons (PAH) molecules. In each
point of the galaxy and for each grain type the appropriate
temperature is computed (either equilibrium $T$ for big grains or
probability distribution of temperature for small grains and PAHs).
The detailed PAH emission spectrum has been updated in Vega et al.
(2005) based on the Li \& Draine (2001) model. (v) The radiative
transfer of starlight through the dust distribution is computed
yielding the emerging SED. The simple stellar population (SSP) library
(Bressan, Granato, \& Silva 1998; Bressan et al. 2002) includes the
effect of the dusty envelopes around AGB stars, and the radio emission
from synchrotron radiation and from ionized gas in HII regions.

The {\gs} parameters that are not provided by {\gal} are summarized in
Table \ref{tab:grasil}, and are described in the following. (i) The
escape time-scale of young stars for the parent MCs is set to $t_{\rm
  esc} = 10^7$ yr. This is intermediate between the values found by
Silva et al. (1998) to well describe the SED of spirals ($\sim$ a few
Myr) and starbursts ($\sim$ a few 10 Myr), and is of the order of the
estimated destruction time scale of MCs by massive stars (e.g. Monaco
et al 2004b). (ii) The gas mass predicted by {\gal} is subdivided
between the dense and diffuse phases, fixing the fraction of gas in
molecular clouds $f_{\rm MC}$ to $0.5$. The resulting SEDs are not
much sensitive to this choice. (iii) The mass of dust is obtained by
the gas mass and the dust to gas mass ratio $\delta$ which is set to
evolve linearly with the metallicity ($\delta=0.45 \; Z$). (iv) The
optical depth of MCs, driving their spectrum, depends on the mass and
radius set for MCs through $\tau \propto \delta \; M_{\rm MC}/r_{\rm
  MC}^2$, with $M_{\rm MC}=10^6$ $M_{\sun}$, $r_{\rm MC} = 16$ pc;
(iv) the bulge and disk scale radii for stars and gas are given by
{\gal}, while the disk scale heights, $h_d^*$ and $h_d^d$ for stars
and dust respectively, are set to $0.1$ the corresponding scale radii.
In addition, the dust grain size distribution and abundances are set
to match the mean Galactic extinction curve and emissivity (as in
Silva et al. 1998 and Vega et al.  2005) and are not varied here.

\begin{table}
\centering \caption{Adopted values for the {\gs} parameters that are not provided by {\gal}.}
\label{tab:grasil}
\begin{tabular}{|lc|}
\hline\hline
Parameters & Values \\
\hline
$t_{\rm esc}$  & $10^7$ yr\\
$f_{\rm MC}$   & 0.5 \\
$M_{\rm MC}/r_{\rm MC}^2$ & $10^6$ M$_{\odot}$/(16 pc)$^2$ \\
$h_d^*$/$r_d^*$  & 0.1 \\
$h_d^d$/$r_d^d$  & 0.1 \\
\hline
\end{tabular}
\end{table}

\subsection{Interfacing {\gal} with {\gs}}
\label{section:interface}

The output of {\gal} consists, for each galaxy, in a time sampling of
the main dynamical variables of the model: for each component (halo,
bulge, disc) the code issues mass, kinetic energy and metal mass of
cold gas; mass, thermal energy and metal mass of hot gas; mass and
metals of stars; average (over the time bin) and punctual (at the end
of the time bin) star formation rate (these are given only for bulge
and disc).  For each galaxy it gives also expelled mass and metals,
black hole mass, cooling radius, bulge and disc radii and velocities,
punctual values of the accretion rate onto the black hole. Star
formation histories are reconstructed following all the exchanges of
stellar matter between galaxies and galaxy components, so that they
refer to the stars contained in the
component at a given time and not to the stars formed in that component.\\

Of this information {\gs} uses the star formation and cold gas
metallicity history for bulge and disk, the mass of gas in the two
components at the required galactic age, and the stars and gas scale
radii at the same age. The time sampling of these quantities is $0.1$
Gyr. {\gs} resamples the star formation histories on a much finer time
grid to optimally account for the short lifetimes of massive stars.
However, the width of the time bin of {\gal} is rather large, so the
number of massive stars can be rather inaccurate.  We then split the
last time bin into a sub-bin of 10 Myr, to which we assign the
punctual value of the star-formation rate at the end of the bin, and a
larger, earlier one of 90 Myr\footnote{As the age of Universe in the
  assumed cosmology is 13.47 Gyrs, the time bin corresponding to $z=0$
  is smaller than 0.1 Gyr; in this case we make the earlier bin
  consistently smaller than 90 Myr.}, to which we assign a star
formation rate such that the integral in the two sub-bins gives the
correct final amount of stars. Using test star-formation histories, we
have verified that this sampling allows to reproduce the magnitudes to
within $<$0.1 mag; indeed, even the strongest star-formation events do
not have associated time-scales much smaller than the time bin.



It is worth mentioning that running the spectro-photometric code on
model galaxies is the main bottleneck of the computation; it is then
necessary to device strategies to optimize the computations
by estimating the minimal number of galaxies needed to have a reliable
result.

\subsection{Luminosity functions, number counts and redshift
  distributions}

\begin{figure}
\centerline{
\includegraphics[width=9cm]{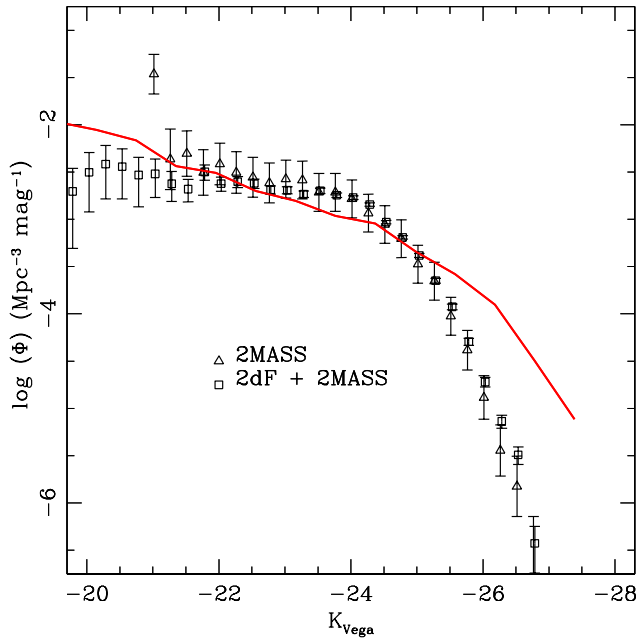}
}
\caption{$K$-band Luminosity Function. Data refer to the observations
  of Cole et al. (2001) and Kochanek et al. (2001). Solid line refers
  to {\gal} prediction.}
\label{fig:klf}
\end{figure}

To simulate a deep field, the computation of galaxy SEDs is performed
on the box at several output times.
As explained in paper I, the sparse-sampling of trees
(section~\ref{section:runs}) results in an over-sampling of small
satellites with respect to central galaxies of similar stellar mass.
We correct for this over-sampling by further sparse-sampling the
satellites as follows. First, we construct from the results of the run
at $z=0$ an average curve of mass of the central galaxy as a function
of DM halo mass; second, we randomly sparse-sample the satellites with
a probability equal to the ratio between the weights $w_{\rm tree}$ of
the tree the satellite belongs to and that of the DM halo whose
central galaxy has on average the same mass as the satellite.  The
inverse of this probability is a new weight, $w_{\rm sat}$.  Central
galaxies are all selected and assigned a unity weight.  If a galaxy is
destroyed by mergers (or tides) then its stellar mass at the
destruction time is used to compute $w_{\rm sat}$.  This procedure
gives a roughly constant number of galaxies in logarithmic intervals
of mass. Because each galaxy is present in many time bins, the number
of selected galaxies is typically very high.  To limit this number we
introduce a third weight $w_{\rm gal}$ as the inverse of a further
sampling factor, equal for all galaxies. The first two samplings (of
merger trees and galaxies) are both computed at $z=0$, but the weights
$w_{\rm tree}$ and $w_{\rm sat}$ assigned to the galaxies are used at
any redshift.  This is done with no loss of generality, as a fair
reconstruction of luminosity functions or number counts only requires
that the weights are used consistently with the sparse-sampling
procedure; in other words, we only need to require that the properties
of a sparsely sampled population are weighted by the inverse of the
sampling probability, whatever the redshift at which the sampling is
performed.

In order to ensure a smooth description of the redshift evolution of
the properties of our simulated galaxies, we use the following
procedure. At every integration time-bin, corresponding to a redshift
$z(t)$, at the end of integration we apply the three sampling
procedures to all the galaxies in the box and select a subsample. It
is worth to notice that the random sparse sampling ensures that a
different subsample of galaxies is considered at each $z(t)$, so that
galaxies at different redshifts are not always the same galaxies seen
at different times. We define a fourth weight $w_{\rm time}$ as the
ratio between the width of the time sampling ($0.1 Gyr$) and the time
interval span by $l_{box}$ at $z(t)$. We also compute the angle
subtended by a square of comoving side $l_{box}$ at $z(t)$. We then
compute the SEDs of model galaxies using {\gs} and we use them to
estimate absolute and apparent magnitudes (both in the Vega and AB
system) and infrared fluxes. We collect these informations in
catalogues.

Luminosity functions, number counts, and redshift distributions of
magnitude-limited samples are then computed by performing weighted
sums over the galaxies using the product of the four weights defined
above.  Luminosity functions at high redshift are computed over the
same redshift intervals as in the observations, while number counts
and redshift distributions of magnitude-limited samples are computed
using galaxies starting form $z \simeq 6$.

\section{Results}
\label{section:results}

\subsection{$K$-band}
\label{section:zzero}

\begin{figure}
\centerline{
\includegraphics[width=9cm]{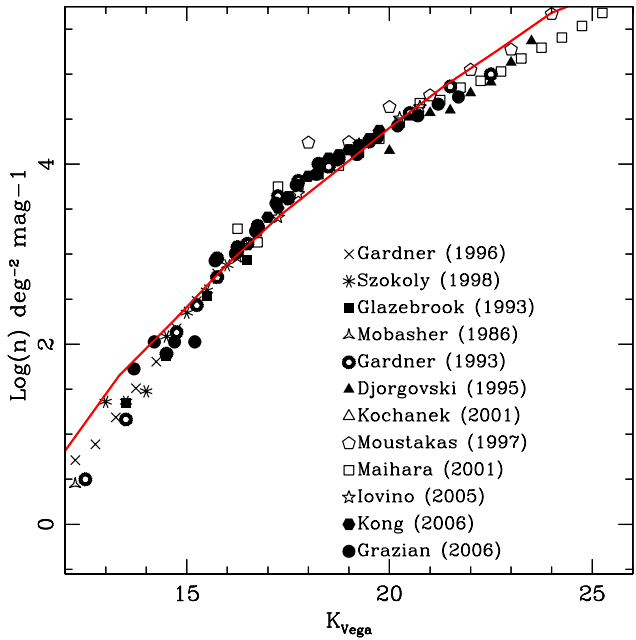}
}
\caption{Source Number Counts in the $K$-band. Data refer to
  observations as listed in the figure.  Solid line refers to {\gal}
  prediction.}
\label{fig:knc}
\end{figure}

The resulting $K$-band LF at $z=0$ (in the Vega system) is compared in
fig.~\ref{fig:klf} to that obtained using the local 2MASS sample
(Jarrett et al. 2000), limited to $K_{Vega}<13.5$ (Kochanek et al.
2001), and the combination of the 2MASS and 2dF samples (Cole et al.
2001). Being the K-band luminosity a good tracer of stellar mass, this
comparison is analogous that of figure~7 of paper I, where the stellar
mass function was compared to that inferred by data (obtained also
with the same 2MASS+2dF sample). In agreement with that result, the
$z=0$ $K$-band LF shows an overestimate of both the bright (massive)
end and the faint (low-mass) end of the LF. In particular, the
overestimate of the high tail is much more relevant here, with respect
to the estimated mass function. We find that this is in part due to a
discrepancy in the adopted $M_*/L_K$ ratios when passing from the
luminosity function to the mass function. The average $M_*/L_K$ ratio
of the {\gal}+{\gs} model galaxies is $\sim 0.7$ M$_\odot/$L$_{K
  \odot}$, while for instance Cole et al. (2001) adopt an higher
average value, $M_*/L_K = 1.32$ M$_\odot/$L$_{K \odot}$, typical of
very old stellar populations (see e.g. figure 24 in Maraston 2005). In
spite of these discrepancies, the model is able to reproduce correctly
the overall normalization of the LF.

Fig.~\ref{fig:knc} shows the predicted $K$-band number counts compared
to data available in the literature (Mobasher et al. 1986; Glazebrook
et al. 1993; Gardner et al. 1993, 1996; Djorgovski et al. 1995;
Moustakas et al. 1997; Szokoly et al. 1998; Kochanek et al. 2001;
Maihara et al. 2001; Iovino et al. 2005; Kong et al. 2006; Grazian et
al. 2006). The models fit well the data in the range $15\la K_{Vega}
\la 22$.  The excess at bright fluxes is a signature of the
overestimate of the LF at the bright end, while the excess at faint
fluxes is commented below.

\begin{figure}
\centerline{
\includegraphics[width=9cm]{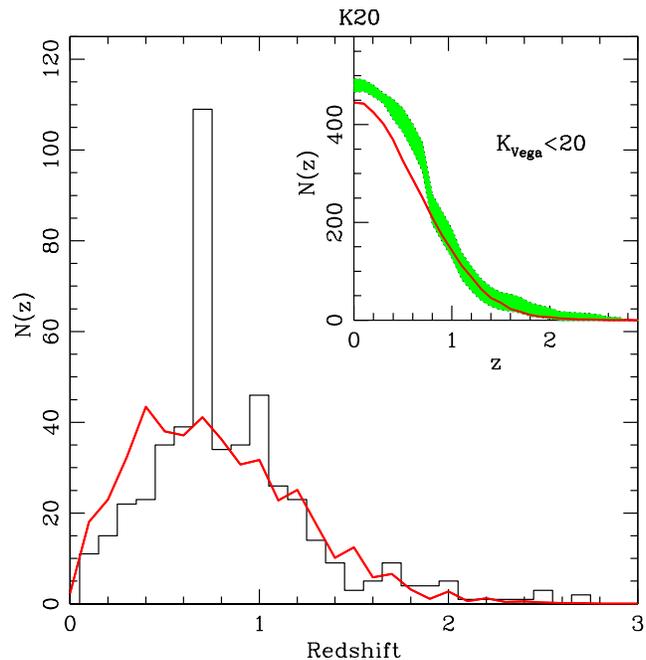}
}
\caption{Redshift source distribution compared to K20 sample
  (histogram, Cimatti et al. 2002a). Solid line refers to {\gal}
  prediction.}
\label{fig:zd_k20}
\end{figure}

\begin{figure*}
  \centerline{ \includegraphics[width=9cm]{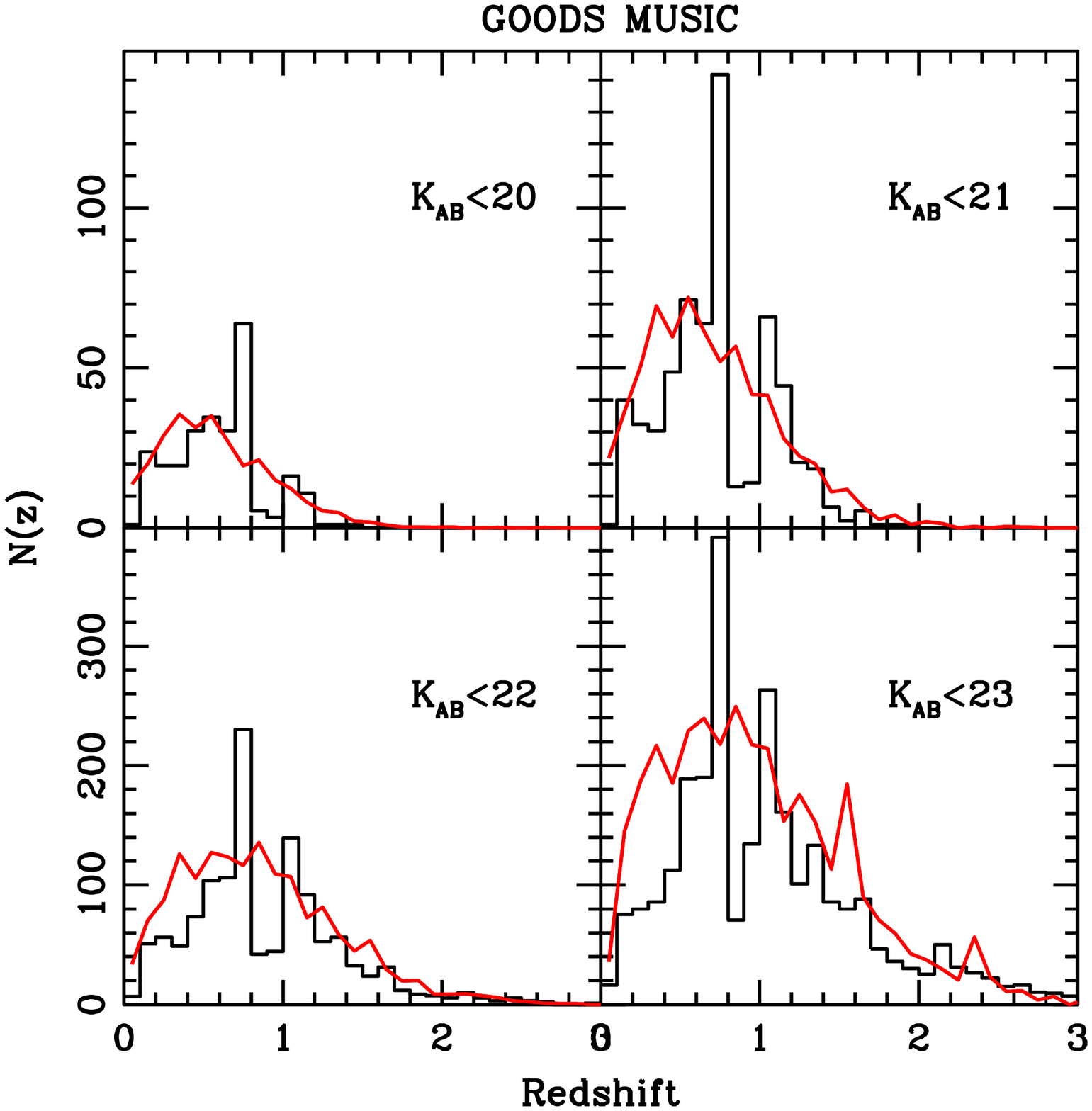}
    \includegraphics[width=9cm]{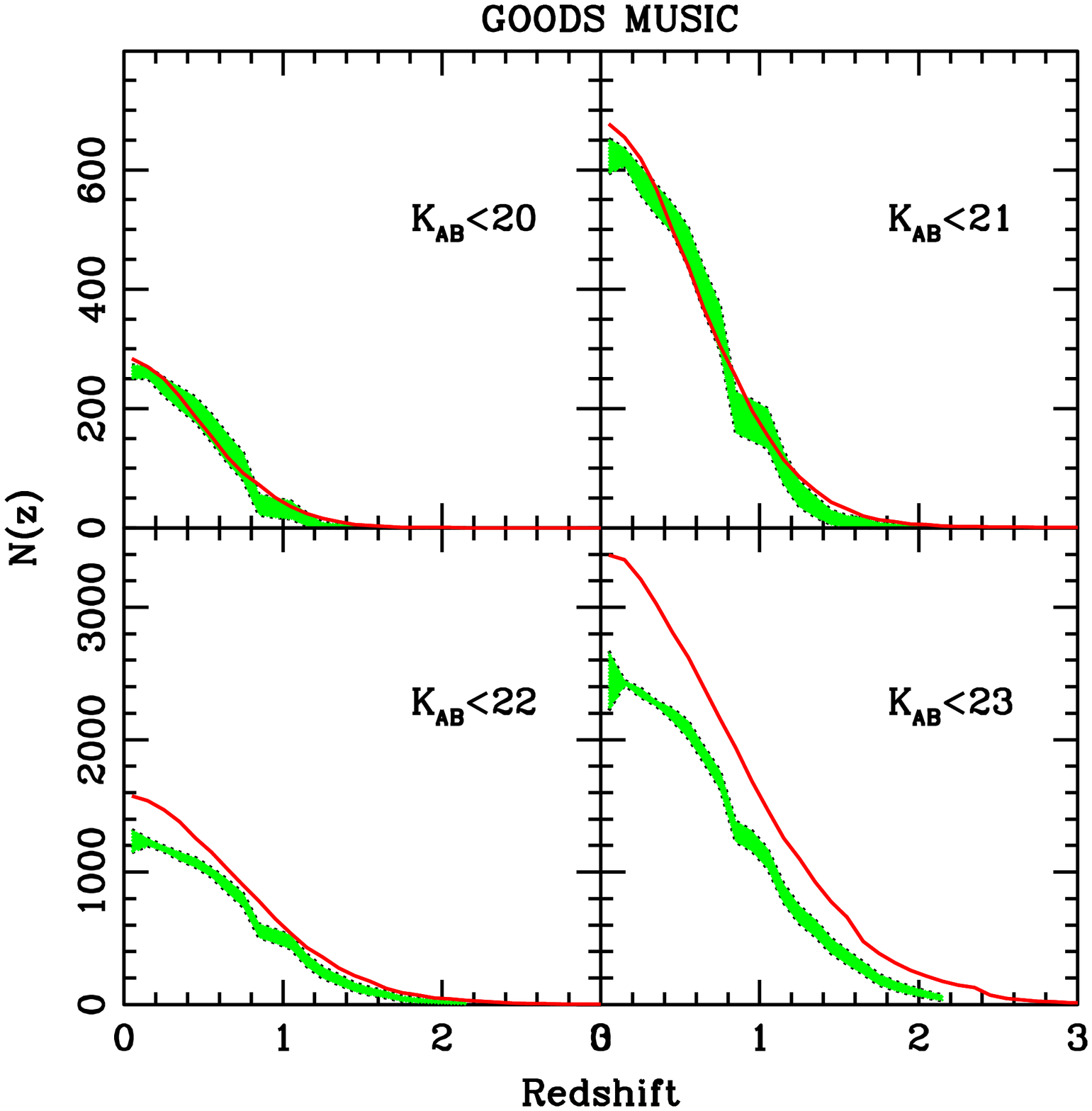} }
  \caption{Redshift source distribution compared to GOODS-MUSIC
    sample (histograms, Fontana et al. 2006). Left Panel: redshift
    distributions.  Right Panel: cumulative distributions. Solid line
    refers to {\gal} predictions.}
\label{fig:zd_GM}
\end{figure*}

\begin{figure}
\centerline{
\includegraphics[width=9cm]{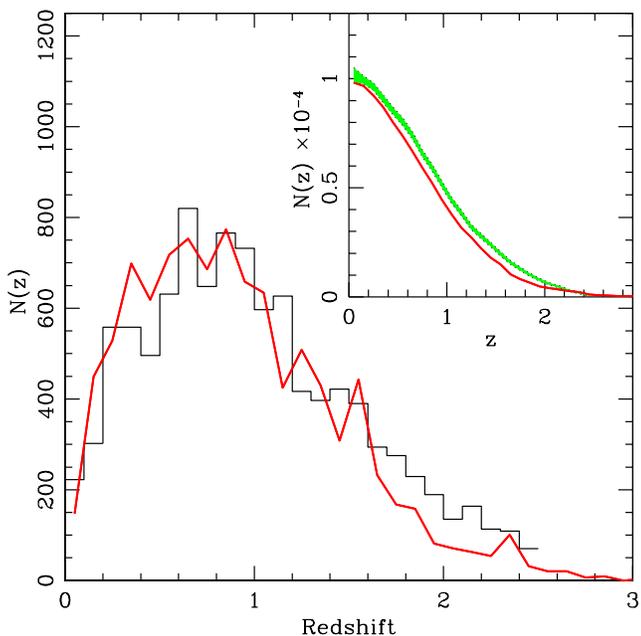}
}
\caption{Redshift source distribution compared to VVDS sample
  (histogram, Pozzetti et al. 2007). Solid line refers to {\gal}
  prediction.}
\label{fig:zd_vvds}
\end{figure}

Number counts do not strongly constrain the model unless redshift
distributions of magnitude-limited samples are available.  We then
compare our results with redshift distributions obtained from the K20
(fig.~\ref{fig:zd_k20}, Vega system), GOODS-MUSIC
(fig.~\ref{fig:zd_GM}, AB magnitudes) and VVDS
(fig.~\ref{fig:zd_vvds}, AB magnitudes) catalogues.  For sake of
clarity we show both the differential and the cumulative
distributions. We compute the error on the cumulative distributions
using a bootstrap technique based on 1000 mock redshift catalogues
drawn using the observed redshift distribution. The agreement of
{\gal} prediction with K20 observations (relative to $K_{AB}<21.84$)
is good: we predict a total number of $435$ objects against $480\pm22$
observed.  The position of the peak of the distribution and the long
tail of galaxies at $z>1.5$ are both recovered; we relate the
disagreement in the cumulative distribution at low redshift to the
known cluster at $z\sim0.7$; some excess at $z\sim0.5$ can be again
connected to the excess of bright galaxies in the K-band LF.
The comparison with the deeper but less wide GOODS-MUSIC sample
(Fig.~\ref{fig:zd_GM}) confirms the good agreement at brighter limit
magnitudes ($301$ predicted vs $262$ observed galaxies with
$K_{AB}<20$; $722$ predicted vs $623$ observed galaxies with
$K_{AB}<21$). At fainter limit magnitudes the comparison highlights an
excess of model galaxies ($1687$ predicted vs $1258$ observed galaxies
with $K_{AB}<22$; $3572$ predicted vs $2603$ observed galaxies with
$K_{AB}<23$). This excess is due not only to the small local excess
noticeable in the $z=0$ LF (Fig.~\ref{fig:klf}), but to a generalized
excess of faint sources at all redshifts.  Fontana et al.  (2006)
showed that the over-prediction of small galaxies ($M\sim10^{10}$
{\msun}) at $z\sim1$ is a common problem of galaxy formation models;
here we see that the problem is probably present since high redshift.
Finally, we also compare {\gal} predictions with the larger VVDS
sample (Fig.~\ref{fig:zd_vvds}, we combined model predictions for the
two sub-samples separately in the same way as observational data): we
obtain again a good agreement with observations ($9822$ predicted
objects against $10160$ observed); except for a mild underestimate,
mostly due to the smaller number of $z\sim1-2$ objects with respect to
observations.

\begin{figure}
  \centerline{
    \includegraphics[width=9cm]{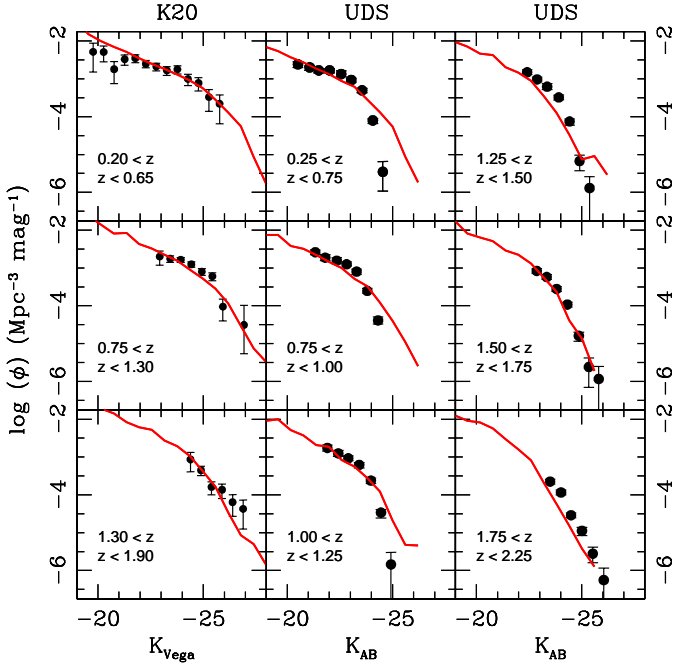} }
  \caption{LF Redshift Evolution. Left panels: K20 sample, data points
    taken from Pozzetti et al. (2003). Mid and Right Panel: UDS
    sample, data points taken from Cirasuolo et al. (2006). Solid line
    refers to {\gal} predictions.}
\label{fig:klf_evo}
\end{figure}

A clearer view is obtained by considering the evolution of the K-band
luminosity function with redshift (Fig.~\ref{fig:klf_evo}).  We
compare our model with the K20 (Pozzetti et al. 2003; Vega system) and
UDS (Cirasuolo et al.  2006; AB magnitudes) data; the K20 estimate is
based on spectroscopic redshifts but the sampled area is small, the
UDS sample covers a larger area of the sky but is based on photometric
redshifts.  The overall agreement between {\gal} and the two datasets
is very good.  We can notice some overestimate of the faint-end at
$z\sim0.5$, which is not as visible as in Fontana et al.  (2006).
Also, the lower redshift bins of the wider UDS sample show that the
excess at the bright end builds up at $z\la1$.

From the analysis of fig.~\ref{fig:klf} to fig.~\ref{fig:klf_evo} we
conclude that {\gal} is able to reproduce the assembly of the bulk of
the stellar mass, which is mostly in place already at $z\sim1-2$;
however, the biggest model galaxies continue growing in mass at $z<1$,
though this growth is partially compensated for by the loss of stars
to the halo by gravitational scattering.  Moreover, the progressive
build-up of smaller objects is not correctly reproduced; small objects
tend to be too many at $z\sim 1$ (and too old at $z\sim0$) in the
model. These findings are in agreement with the analysis presented in
Fontana et al. (2006).

\subsection{850$\mu$m band}
\label{section:scuba}

\begin{figure}
\centerline{
\includegraphics[width=9cm]{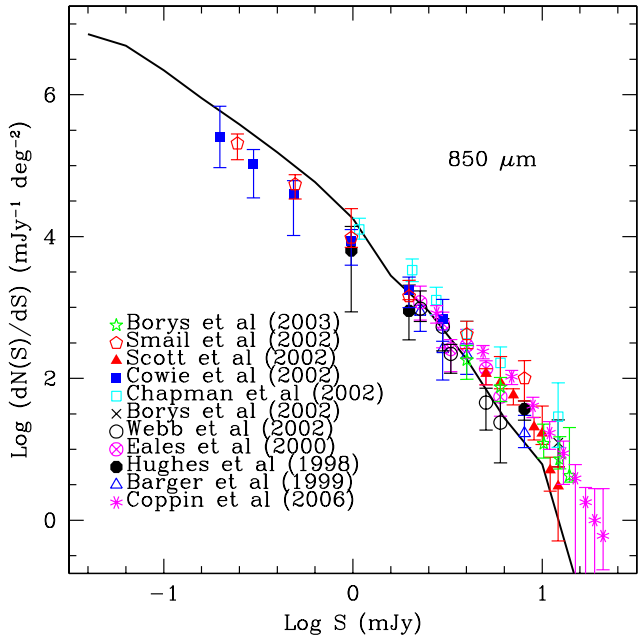}
}
\caption{$850 \mu$m Source Number Counts. Data refer to observations
  as listed in the figure. Solid line refers to {\gal} prediction.}
\label{fig:scuba}
\end{figure}

\begin{figure}
\centerline{
\includegraphics[width=9cm]{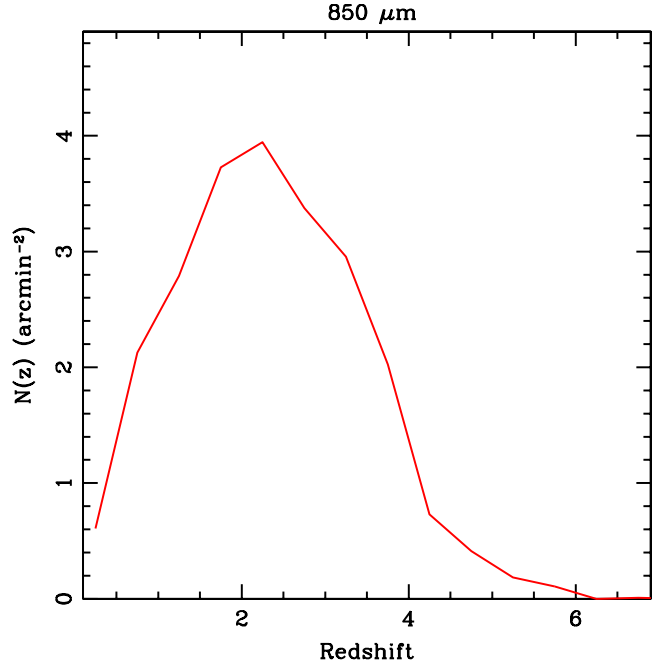}
}
\caption{Predicted Redshift Distribution for $850 \mu$m sources with $S > 0.2 mJy$.}
\label{fig:zd_scuba}
\end{figure}

We now test whether the stars visible at $z<2$ in the K-band were
formed in much smaller chunks that were later assembled together, or
in big massive starbursts at $z\ga 2$.  This is best tested by
comparing the model to number counts at 850$\mu$m, where massive
starbursts at $1\la z \la 5$ are easily observed.
Fig.~\ref{fig:scuba} shows our prediction compared to SCUBA
observations (Hughes et al.  1998; Barger et al. 1999; Eales et al.
2000; Webb et al. 2002; Borys et al. 2002, 2003; Chapman et al.  2002;
Cowie et al. 2002; Scott et al. 2002; Smail et al. 2002; Coppin et al
2006; Scott et al 2006). In order to avoid contamination due to
low-redshift sources, we consider here a subsample with $z>0.5$. We
stress that this prediction is done assuming a standard Salpeter IMF.
Fig.~\ref{fig:zd_scuba} gives the redshift distribution of the objects
with flux $>0.2$ mJy, which corresponds to the faintest point in
fig.~\ref{fig:scuba}. It is peaked at $z\sim2.0$, in qualitatively
agreement with the observations of Chapman et al.  (2003,2005) of the
brightest SCUBA sources (redshift distribution peaked at $z\sim2.4$
with a quartile range of $1.9 < z < 2.8$). The model reproduces well
the data, with a possible modest overestimate (underestimate) at faint
(bright) fluxes. However, a deeper analysis of the predictions shows
that the population of brightest objects ($>5 mJy$) at $z>2$ (Chapman
et al., 2005; Aretxaga et al., 2007), is missing: we are able to
reproduce the bulk of the heavily star-forming population at $z\sim2$,
but not the most luminous objects.

\section{Discussion}
\label{section:discussion}

\begin{figure*}
\centerline{
\includegraphics[width=9cm]{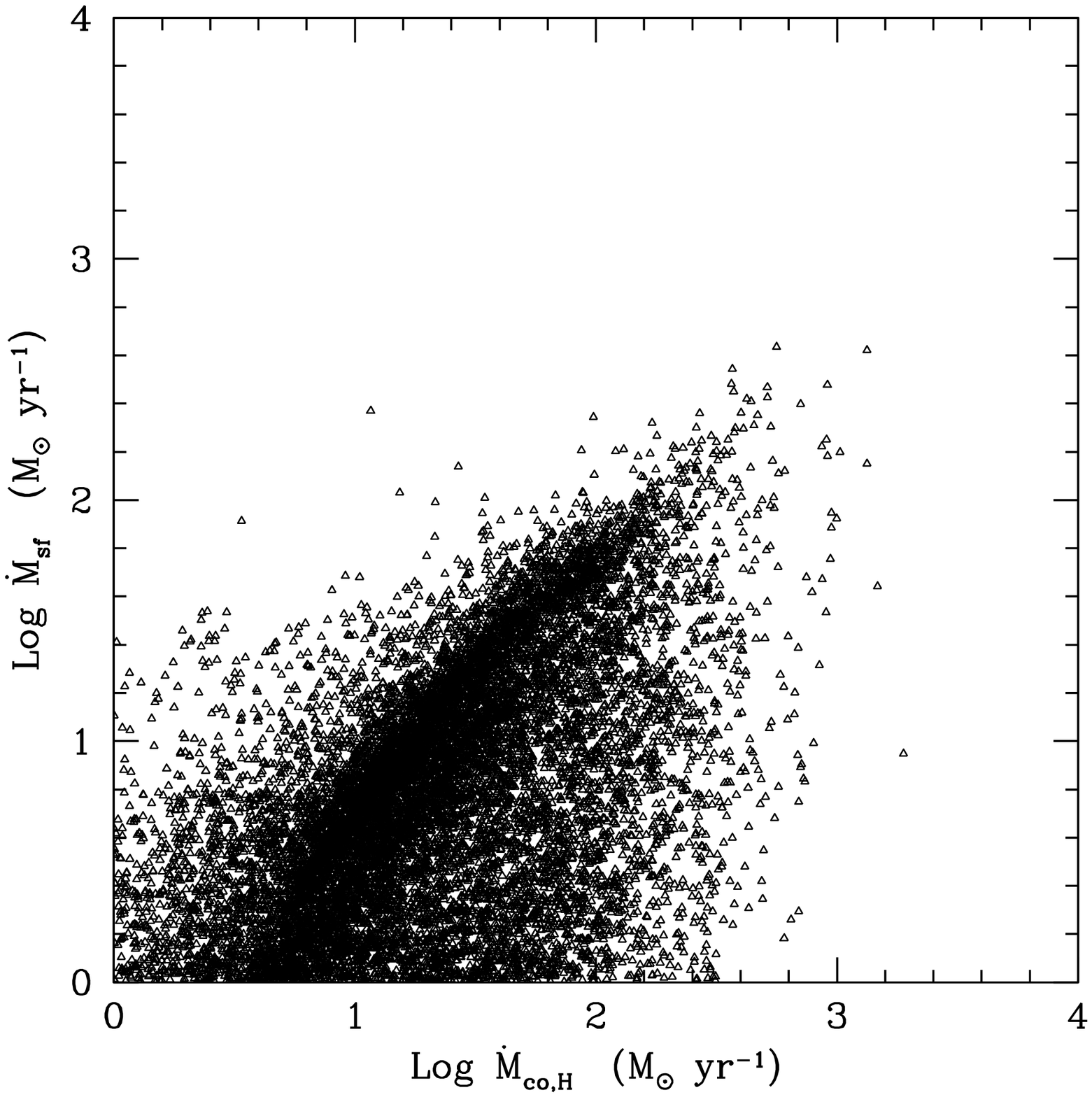}
\includegraphics[width=9cm]{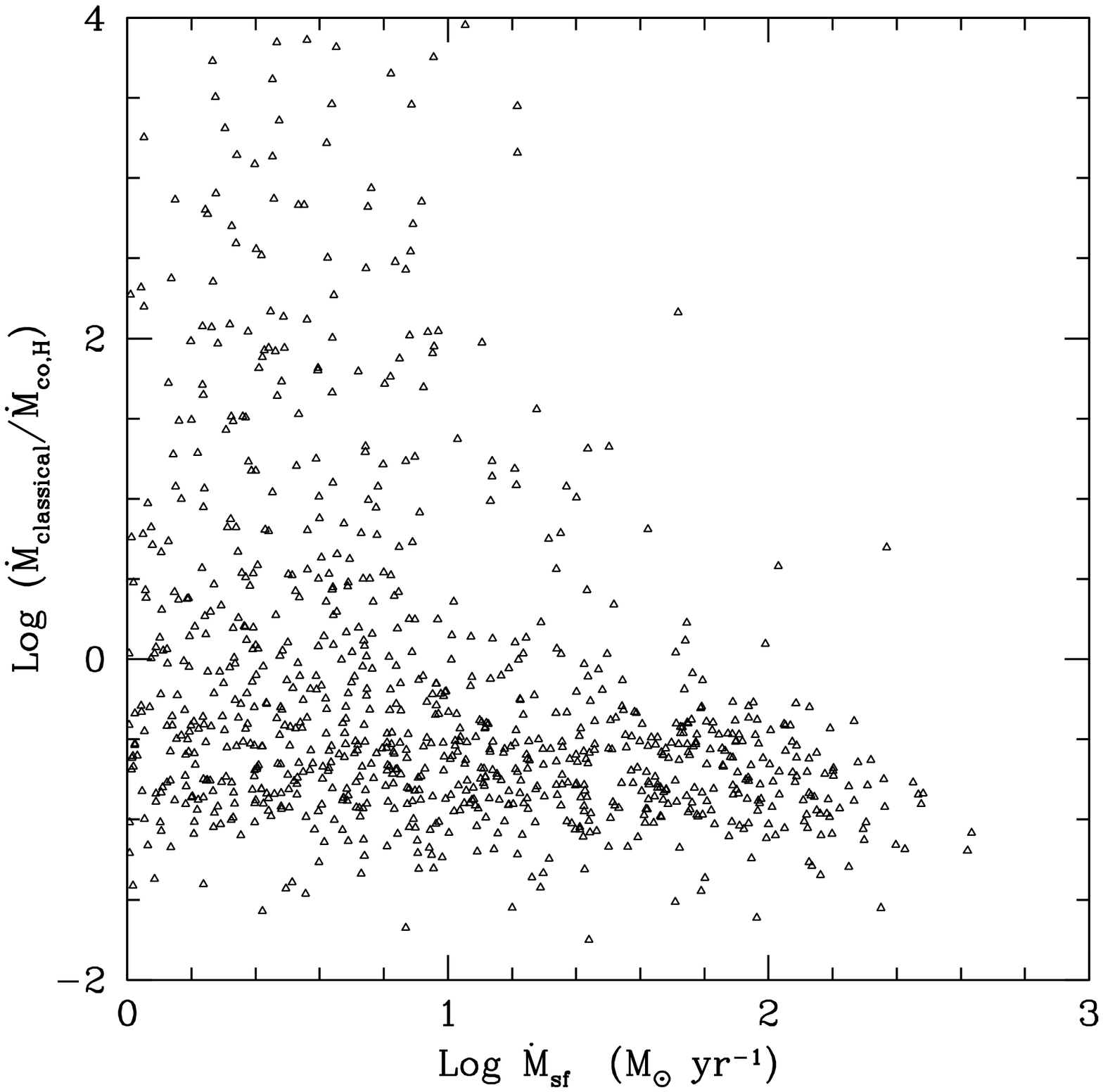}
}
\caption{Left panel: star formation rate (bulge$+$disc) versus cooling
  rate for central galaxies at $z=2.5$.  Right panel: ratio of the
  cooling flows as predicted by the classical model and as used by
  {\sc morgana} versus total star formation rate for the same galaxies
  as above.  In the right panel many points, especially at low
  star-formation, lie out of the plot because the classical model
  predicts that all the gas has cooled, while {\sc morgana} retains
  some gas through feedback and cosmological infall.}
\label{fig:cc}
\end{figure*}

Sub-mm counts have longly been an elusive piece of evidence to fit for
semi-analytic models of galaxy formation (e.g. Guiderdoni et al. 1998;
Devriendt \& Guiderdoni 2000; Granato et al.  2000; but see Baugh et
al. 2005). Conversely, in the long tuning and debugging process of the
{\gal} code, we have never found any difficulty in fitting the bulk of
number counts using a conservative choice for the IMF.  The difference
can be ascribed to the different cooling model that {\gal} is
implementing.  In a forthcoming paper (Viola et al. 2007) we compare
analytic cooling models to N-body+hydro simulations in the simplest
case of an isolated halo with its hot gas component initially in
hydrostatic equilibrium, and no feedback from star formation or AGN.
The classical model of cooling, based on the computation of a cooling
radius (White \& Frank 1991) is found to underestimate the amount of
cooled gas, especially in the first stages of cooling, while the model
implemented in {\gal} and briefly described in
Section~\ref{section:morgana} gives a very good fit.  All the analytic
models of cooling tend to give similar results at later times.  The
use of a better cooling model accelerates the accumulation of cold gas
in the halos, giving rise to stronger starbursts. The typical mass of
the halos that contain the biggest starbursts is high enough to ensure
the formation of a hot halo component through shocks (see, e.g., Keres
et al. 2005), so that the results based on hot gas halos in
hydrostatic equilibrium apply to this case.

To support this interpretation, we run again the model, computing at
each halo major merger the gas profile (as described in
Section~\ref{section:morgana}) and from it the cooling time $t_{\rm
  cool}(r)$ as a function of radius.  The classical cooling radius
$r_{\rm C}(t)$ is the inverse of the function $t_{\rm cool}(r)$.  The
cooling flow of the classical cooling model is then computed as
$\dot{M}_{\rm classical} = 4\pi r_{\rm C}^2 \rho_g(r_{\rm C}) dr_{\rm
  C}/dt$.  It is not possible to use the classical cooling flow
directly in the model, because the cooling model includes the
injection of energy by feedback, which would then be absent.  Besides,
a generalization of the model to include classical cooling would
require deep changes in the code and a re-calibration of the
parameters to reproduce local observables, which is well beyond the
interest of the paper.  We then simply compare the classical cooling
flow with that actually used in the {\gal} model.  Firstly, in
figure~\ref{fig:cc}, left panel, we show {\gal} cooling flow versus
star formation for all the central galaxies present in the box at
$z=2.5$.  The stronger starbursts are associated with the stronger
cooling flows\footnote{A few massive starburst are associated with no
  cooling flow and then lie outside this relation; these are mergers
  where feedback has already quenched cooling, or where cooling has
  already deposited most mass in the galaxy}.  Second, we show for the
same halos (right panel) the ratio of classical and {\gal} cooling
flows versus star formation.  For massive starbursts ($\dot{M}_{\rm
  sf}>100$ M$_\odot$ yr$^{-1}$) the classical model predicts cooling
flows a factor of 5-10 lower, which would presumably result in
correspondingly lower star formation rates.  The relation changes at
smaller star formation rates, where feedback has been able to suppress
the {\gal} cooling flow in many cases\footnote{Many points at low
  star-formation lie out of this plot because the classical model
  predicts no cooling, while in {\gal} gas is still present due to
  feedback and cosmological infall.}. The success of {\gal} should
then be ascribed to the different modeling of cooling.

\begin{figure*}
\centerline{
\includegraphics[width=9cm]{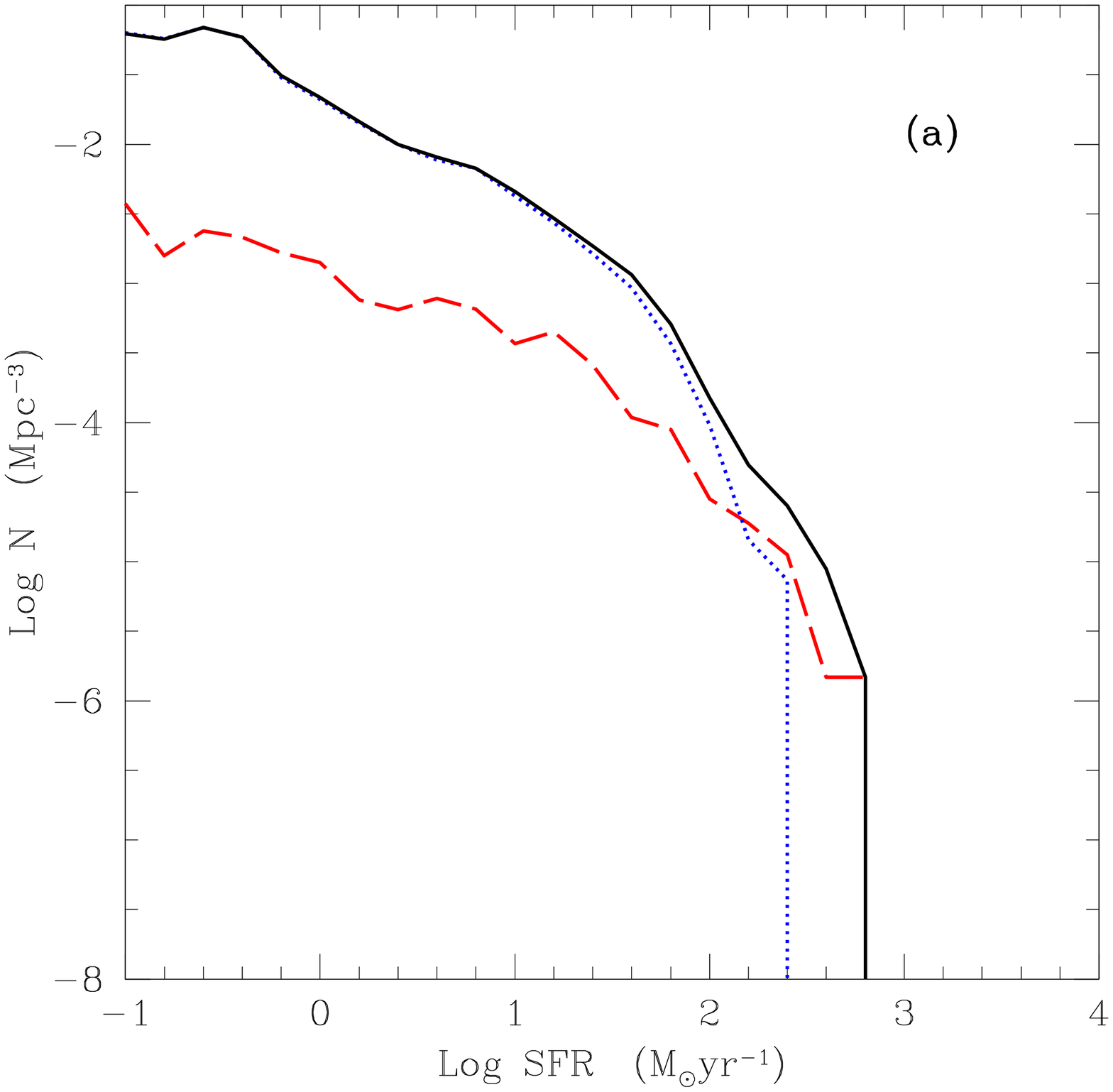}
\includegraphics[width=9cm]{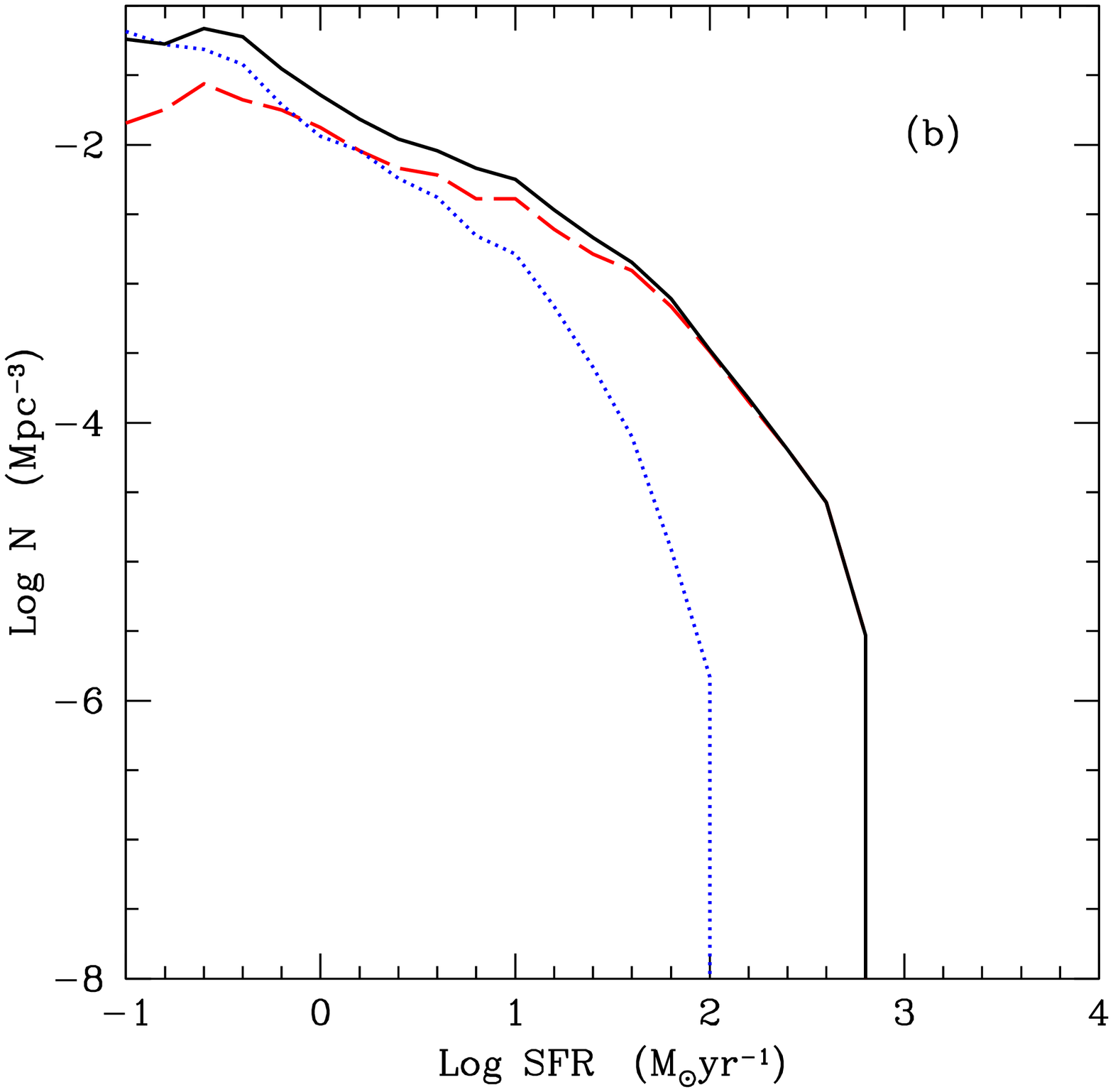}
}
\caption{Star-formation rate function at $z=2.5$ for discs (blue
  dotted line), bulges/mergers (red dashed line) and total (black
  continuous line).  The star formation in panel (a) is the punctual
  value at the end of the integration over the time bin, in (b) is the
  average in the bin (of width 0.1 Gyr) relative to the stars
  contained in bulges and discs at the end of the integration.}
\label{fig:bulges}
\end{figure*}

With our model we can also investigate the physical conditions of
SCUBA sources.  In fig.~\ref{fig:bulges}a we show the star-formation
function of galaxies at $z=2.5$, divided into bulges (mergers) and
discs. In this case, we use the punctual value of the star formation
rate computed at the end of the time-bin.  Interestingly, mergers
dominate only at the highest star formation rates, $\ga200$ \msunyr.
This implies that most of our starbursts are triggered by
cooling/infall more than by mergers, in line with the findings of
violently star-forming discs at $z\sim 2$ (Genzel et al. 2006).
However, this does not imply that massive starbursts are associated
with spiral galaxies.  In Fig.~\ref{fig:bulges}b the star formation
rate is computed as the total amount of stars formed in the 0.1 Gyr
time bin and eventually found (at the end of the time bin) in a bulge
or disc component, divided by the width of the time bin.  This time
bulges clearly dominate the star-formation function.  This shows that
such massively star-forming discs merge into bulges (disc
instabilities are not relevant at this redshift) in less than 0.1 Gyr.
Such short time-scales guarantee high levels of $\alpha$-enhancement,
so the stellar populations formed in these starbursts will be typical
of early-type galaxies.  We conclude that the cooling/infall
domination of high-redshift starbursts does not change the conclusion
that these events are responsible for the formation of the stars found
today in elliptical galaxies.

The overall agreement between model and data is in line with recent
results by De Lucia et al. (2006), Bower et al. (2006), Croton et al.
(2006) and Kitsbichler \& White (2006).  However, there are three main
points of disagreement between our model and the data. First, the
evolution of the most massive galaxies at $z<1$ is too strong,
resulting in an excess of bright galaxies in the K-band.  It is easy
to absorb this excess at $z=0$ by tuning the parameters, but this
would be done at the expense of a poor fit of the stellar mass
function and $K$-band LF at $z\sim1$.  As demonstrated by Monaco et
al. (2006), this evolution is driven by galaxy mergers, so no feedback
recipe can solve it.  Our implementation of the suggestion of Monaco
et al.  (2006) to scatter 40\% stars to the diffuse component of
galaxy clusters at each merger goes in the right direction but is
largely insufficient to suppress this evolution.  A much higher
scattered fraction, like $\sim$80\% of the stellar mass of the
satellite, would give a better results (as argued also by Conroy et
al. 2007 and Renzini 2007).  However, such an extreme value applied to
all mergers and all redshift does not allow to reproduce the $\sim1$\%
fraction of diffuse stars in the Milky Way together with the
$\sim10$\% fraction in Virgo and the $\sim40$\% fraction in massive
clusters.  We conclude that, while the mechanism is promising and
deserves further attention, an easy and straightforward implementation
does not work; this mechanism should be selectively efficient in the
most massive DM halos at low redshift.

Second, the model is not able to reproduce the most luminous SCUBA
galaxies at redshift $z\sim2$, which correspond to the strongest
starbursts. This also hints to an insufficient ``downsizing'' in our
predicted galaxy population. In the most massive galaxies the fraction
formed at high redshift in episode of intense star formation is still
too small, and the fraction of stars accreted at later times is still
too large, with respect to the observed trends. Given the success in
reproducing the bulk of the SCUBA population, a detailed investigation
on the predictions of the cooling/infall model in the more massive
halos is promising. Also the choice of a less conservative IMF, i.e.
the Kroupa IMF, can help to solve the problem, thanks to the larger
number of massive stars predicted at each stellar generation.

\begin{figure*}
\centerline{
\includegraphics[width=9cm]{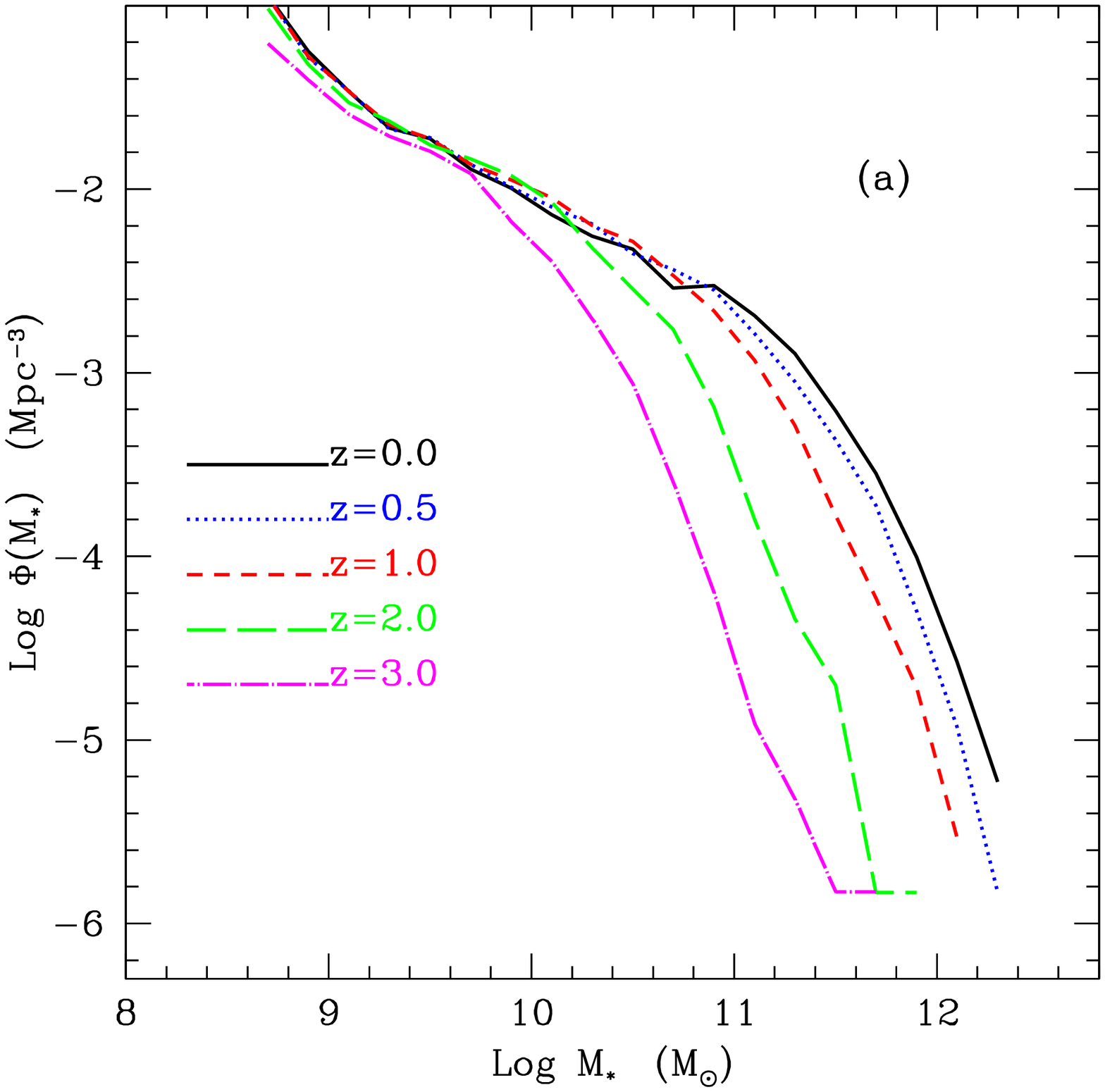}
\includegraphics[width=9cm]{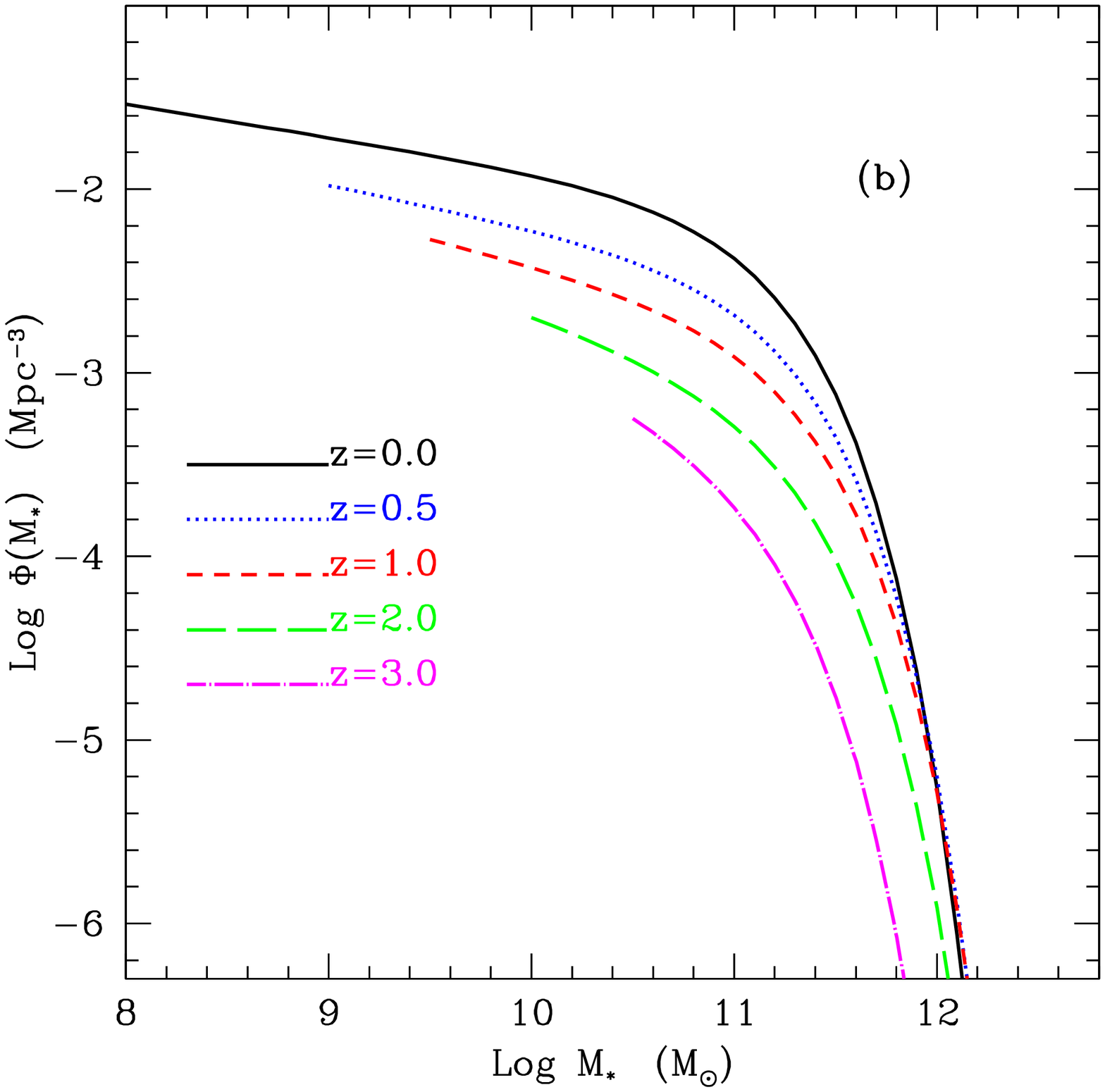}
}
\caption{Evolution of the stellar mass function according to {\gal}
  (panel a) and GOODS-MUSIC (panel b).  The latter mass functions are
  drawn roughly in the same mass interval where data are available.}
\label{fig:down}
\end{figure*}

Third, we confirm the finding of Fontana et al. (2006) that our model
produces too many faint objects at $z\sim1$; in other words, our
smaller galaxies are too old.  Coupled with the excessive evolution of
the most massive galaxies, this finding shows that, while reproducing
the average build-up of galaxies, the observed trend of less massive
galaxies being on average younger is not properly reproduced by
{\gal}. Fontana et al. (2006) have shown very convincingly that this
problem is shared by many galaxy formation codes, either numerical or
semi-analytic, so our results can be considered as the typical outcome
of the $\Lambda$CDM cosmogony, given the present understanding of the
physics of galaxy formation.  Figure~\ref{fig:down} shows a comparison
of the evolution of the stellar mass function as inferred by the
GOODS-MUSIC sample and as predicted by {\gal}.  Clearly, the
``downsizing'' trend of more massive galaxies evolving very slowly at
$z<1$, while the population of less massive galaxies builds up, is not
reproduced.  While the evolution of the high-mass end is driven by
mergers, a delay in the build-up of faint galaxies should be caused by
some source of feedback.  

\section{Conclusions}
\label{section:conclusions}

This paper is the third of a series devoted to presenting {\gal}.  We
have demonstrated that the model is able to follow the build-up of the
bulk of stars, more precisely to reproduce the early assembly and
late, almost-passive evolution of massive galaxies.  To this aim we
have combined {\gal} with {\gs} and compared predictions with
observation in the submm (the 850$\mu$m channel), especially sensitive
to the strongest and most obscured episodes of star formation, and in
the NIR (especially in the $K$-band), most sensitive to the stellar
mass.  Overall consistency between model and observations has been
obtained using SCUBA counts at
850$\mu$m, 
number counts in the $K$-band, redshift distribution of $K$-limited
galaxy samples, and redshift evolution of the $K$-band LF.
The importance of this result is strengthened by the use of a standard
Salpeter IMF (e.g. a very conservative choice) along the whole
redshift interval.  We ascribe this agreement mostly to our improved
model for cooling/infall, that correctly reproduces the results
controlled numerical experiments.


We predict that most star formation at high redshift is not stimulated
by starbursts but is due to the strong cooling/infall flows that take
place at early times. This does not imply that such discs are close
analogues of local spiral galaxies, as their gas surface density is
very high, their star formation rate is typical of starburst galaxies
and most of them are expected to merge within less than 0.1 Gyr. We
then predict that gas-rich discs, characterized by star-formation
rates up to $100-200$ {\msunyr}, should be very abundant at $z\sim2$,
in line with the observations of Genzel et al. (2006).



Despite these successes, our model does not reproduce the
``downsizing'' trend of a modest evolution of the most massive
galaxies accompanied by a build-up of small galaxies at $z\la1$.  We
propose that a solution of this discrepancy requires at least two
mechanisms, because the evolution of the bright end is driven by
inevitable galaxy mergers, while star formation in less massive
galaxies is sensitive to feedback.  The solution proposed by Monaco et
al. (2006) for slowing the evolution of massive galaxies requires an
implementation of scattering of stars to the diffuse component that is
strongly dependent on DM halo mass and/or redshift. On the other hand,
a solution of the overabundance of $10^{11}$ {\msun} galaxies at
$z\sim1$, common to most galaxy formation models (Fontana et al.
2006), calls for some unknown source of feedback.

\section*{Acknowledgments}
We thank Stefano Cristiani, Andrea Cimatti, Gianluigi Granato, Adriano
Fontana, Alvio Renzini, Carlos Frenk, Cedric Lacey, Rachel Somerville
and Eric Bell for many enlightening discussions; we also thank Lucia
Pozzetti for providing the VVDS redshift distribution. PM thanks the
ICC of Durham for its hospitality. Calculations were carried out both
at the ``Centro Interuniversitario del Nord-Est per il Calcolo
Elettronico'' (CINECA, Bologna) with CPU time assigned under
University of Trieste/CINECA grants, and at the PIA cluster of the
Max-Planck-Institut f\"ur Astronomie at the Rechenzentrum Garching.

{}

\bsp

\label{lastpage}
\end{document}